\documentclass[aps,prl,superscriptaddress,twocolumn,showpacs]{revtex4-1}
\usepackage{color}

\usepackage{graphicx}
\usepackage[ansinew]{inputenc}
\usepackage{epstopdf}

\begin{document}

\title{Topological Hopf and chain link semimetal states and their application to Co$_2$MnGa (Theory and Materials Prediction)}
\author{Guoqing Chang$^*$\footnote[0]{*These authors contributed equally to this work.}}
\affiliation{Centre for Advanced 2D Materials and Graphene Research Centre National University of Singapore, 6 Science Drive 2, Singapore 117546}\affiliation{Department of Physics, National University of Singapore, 2 Science Drive 3, Singapore 117542}

\author{Su-Yang Xu$^*$}\affiliation {Laboratory for Topological Quantum Matter and Spectroscopy (B7), Department of Physics, Princeton University, Princeton, New Jersey 08544, USA}
\author{Xiaoting Zhou$^*$}\affiliation{Centre for Advanced 2D Materials and Graphene Research Centre National University of Singapore, 6 Science Drive 2, Singapore 117546}\affiliation{Department of Physics, National University of Singapore, 2 Science Drive 3, Singapore 117542}

\author{Shin-Ming Huang}
\affiliation{Department of Physics, National Sun Yat-sen University, Kaohsiung 804, Taiwan}

\author{Bahadur Singh}\affiliation{Centre for Advanced 2D Materials and Graphene Research Centre National University of Singapore, 6 Science Drive 2, Singapore 117546}\affiliation{Department of Physics, National University of Singapore, 2 Science Drive 3, Singapore 117542}
	
\author{Baokai Wang}
\affiliation{Department of Physics, Northeastern University, Boston, Massachusetts 02115, USA}

\author{Ilya Belopolski}\affiliation {Laboratory for Topological Quantum Matter and Spectroscopy (B7), Department of Physics, Princeton University, Princeton, New Jersey 08544, USA}
\author{Jiaxin Yin}\affiliation {Laboratory for Topological Quantum Matter and Spectroscopy (B7), Department of Physics, Princeton University, Princeton, New Jersey 08544, USA}
%\author{Hao Zheng}\affiliation {Laboratory for Topological Quantum Matter and Spectroscopy (B7), Department of Physics, Princeton University, Princeton, New Jersey 08544, USA}

%\author{Daniel S. Sanchez}\affiliation {Laboratory for Topological Quantum Matter and Spectroscopy (B7), Department of Physics, Princeton University, Princeton, New Jersey 08544, USA}
\author{Songtian Zhang}\affiliation {Laboratory for Topological Quantum Matter and Spectroscopy (B7), Department of Physics, Princeton University, Princeton, New Jersey 08544, USA}

\author{Arun Bansil}
\affiliation{Department of Physics, Northeastern University, Boston, Massachusetts 02115, USA}

%\author{Han Hsu}
%\affiliation{Department of Physics, National Central University, Jhongli City, Taoyuan 32001, Taiwan}

\author{Hsin Lin$^{\dag}$}
\affiliation{Centre for Advanced 2D Materials and Graphene Research Centre National University of Singapore, 6 Science Drive 2, Singapore 117546}
\affiliation{Department of Physics, National University of Singapore, 2 Science Drive 3, Singapore 117542}

\author{M. Zahid Hasan $^{\dag}$\footnote[0]{$^{\dag}$Corresponding authors (emails): suyangxu@princeton.edu, nilnish@gmail.com, mzhasan@princeton.edu }}\affiliation{Laboratory for Topological Quantum Matter and Spectroscopy (B7), Department of Physics, Princeton University, Princeton, New Jersey 08544, USA}
\affiliation{Lawrence Berkeley National Laboratory, Berkeley, California 94720, USA}

\begin{abstract}
Topological semimetals can be classified by the connectivity and dimensionality of the band crossing in momentum space. The band crossings of a Dirac, Weyl, or an unconventional fermion semimetal are zero-dimensional (0D) points, whereas the band crossings of a nodal-line semimetal are one-dimensional (1D) closed loops. Here we propose that the presence of perpendicular crystalline mirror planes can  protect three-dimensional (3D) band crossings characterized by nontrivial links such as a Hopf link or a coupled-chain, giving rise to a variety of new types of topological semimetals. We show that the nontrivial winding number protects topological surface states distinct from those in previously known topological semimetals with a vanishing spin-orbit interaction. We also show that these nontrivial links can be engineered by tuning the mirror eigenvalues associated with the perpendicular mirror planes. Using first-principles band structure calculations, we predict the ferromagnetic full Heusler compound Co$_2$MnGa as a candidate. Both Hopf link and chain-like bulk band crossings and unconventional topological surface states are identified.\end{abstract}

\maketitle

Since the discovery of Dirac and Weyl semimetals \cite{Weyl, Herring, Volovik2003, Murakami2007, Wan2011, Burkov2011, Hsin_RMP, Na3Bi, Na3Bi_3, Na3Bi_2, Cd3As2_2, Cd3As2_3, Cd3As2_4, Huang2015, Weng2015, Hasan_TaAs, TaAs_Ding, MIT_Weyl,Rev1, Rev2}, topological semimetals have emerged as an active frontier in condensed matter physics. Their unique topological properties are predicted to give rise to a wide range of exotic transport and optical phenomena \cite{Hasan2010, Qi2011, nielsen1983adler, Nonlocal, Fermi arc_1, Fermi arc_2, Inti, PC_Weyl_Tanaka, PC_Weyl_Nagaosa, PC_Weyl_Chris, PC_Weyl_Moore, Moore2, Weyl_Floquet, Burkov, Hosur, Pallab, Pesin}. By considering various rotational and mirror symmetries in both symmorphic and non-symmorphic contexts, researchers have predicted nodal-line semimetals \cite{Burkov_NL}, higher charge double Weyl states \cite{Chen, SrSi2}, eightfold-degenerate double Dirac fermions \cite{Double Dirac}, non-symmorphic nodal-chain metals \cite{Chain}, the single threefold- and doubled sixfold-degenerate spin-1 Weyl points \cite{New Fermion}, nexus fermions \cite{Nexus, trip1, trip2}, Kramers Weyl fermions \cite{Kramers Weyl}, and magnetic Dirac semimetals \cite{AFM, AFM_2, AFM_3}. Despite this diversity, topological semimetals can be further classified and characterized by the dimensionality of their band crossings in the bulk Brillouin zone (BZ). In a Dirac/Weyl semimetal or an unconventional (higher-fold degenerate) fermion semimetal \cite{Nexus, trip1, trip2, Double Dirac, New Fermion}, the conduction and valence bands cross at discrete points in the BZ. Therefore, the dimension of their band crossings is 0D. In a nodal-line semimetal \cite{Burkov_NL}, the conduction and valence bands touch along a closed loop, thus the dimension of its band crossing is 1D. In this letter, we propose a number of previously unidentified topological semimetals and identify a candidate material class for the experimental realization. They feature \textit{3D} band crossings characterized by nontrivial links such as a Hopf link or a coupled-chain enabled by perpendicular mirror planes. The Hopf link, which consists of two rings that pass through the center of each other, represents the simplest topologically nontrivial link. While originally studied in mathematics and other areas, recently, researchers have applied this concept into topological physics in order to construct novel topological insulators and superconductors \cite{Hopf_in1, Hopf_in2, Hopf_in3}, although the role of Hopf link is distinctly different from what is considered here. Here we apply this idea in metals and show that the concept of Hopf link can give rise to exotic, previously uncharacterized topological semimetals. We show that these nontrivial links can be identified by engineering the mirror eigenvalues of the conduction and valence bands on the two perpendicular mirror planes. We theoretically identify the ferromagnetic full Heusler compound Co$_2$MnGa(Co$_2$MnAl) as a candidate to experimentally realize this novel state of matter.

\begin{figure}
\includegraphics[width=88mm]{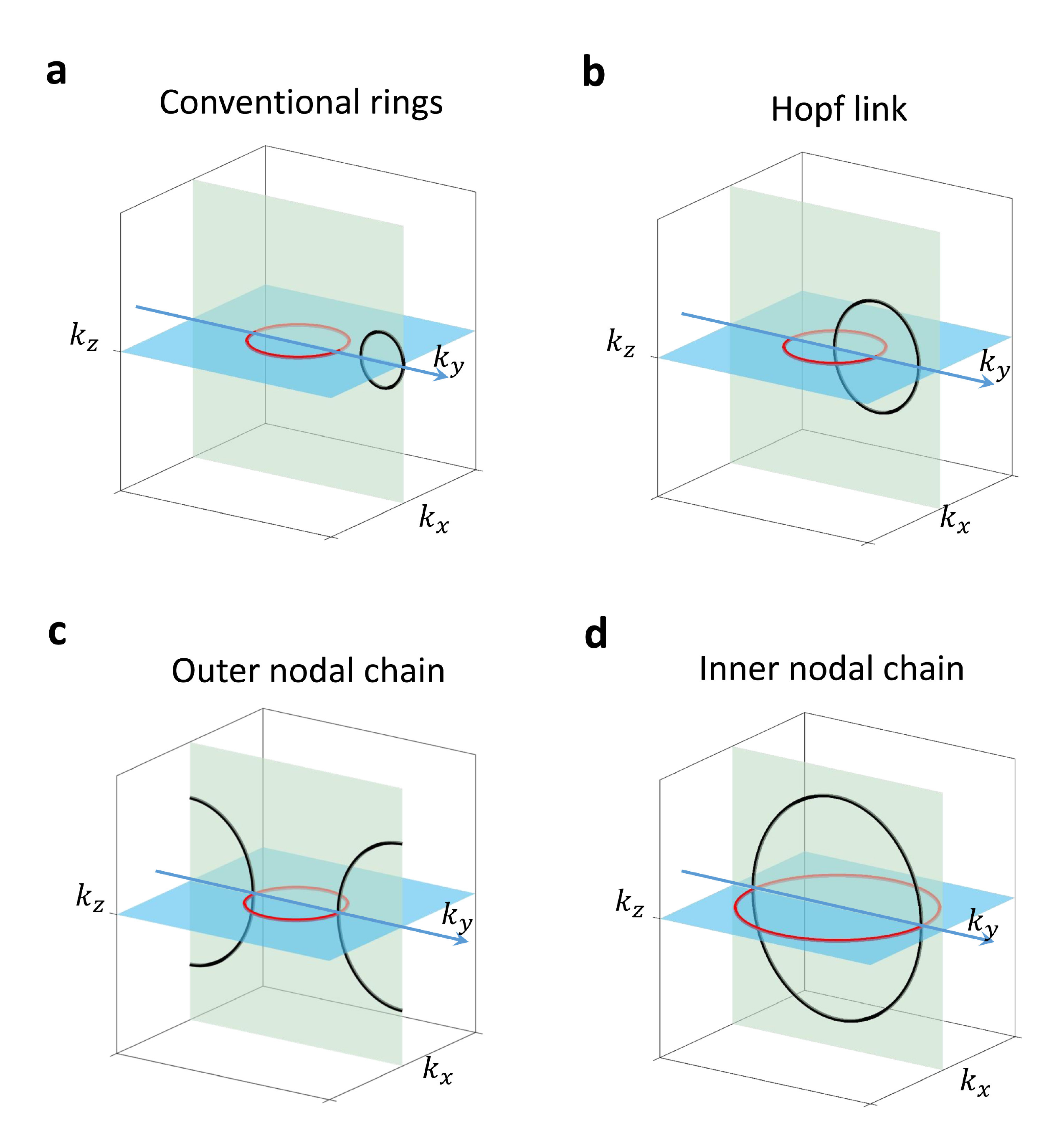}
\caption{\textbf{A variety of new topological semimetals characterized by 3D nontrivial links.} ({a}) Two independent nodal-liness protected by two perpendicular mirror planes. ({b}) When each nodal-line passing through the center of the other, the two nodal-lines form a nontrivial structure, a Hopf link. A Hopf link is a 3D structure which can not be mapped to any 2D surface. ({c, d}) When the two nodal-lines touch at a point, they form a chain. When the two nodal-lines are on opposite (same) sides of the touching point, they form a outer (inner) nodal chain.}
\label{Fig1}
\end{figure}

\begin{figure}
\includegraphics[width=88mm]{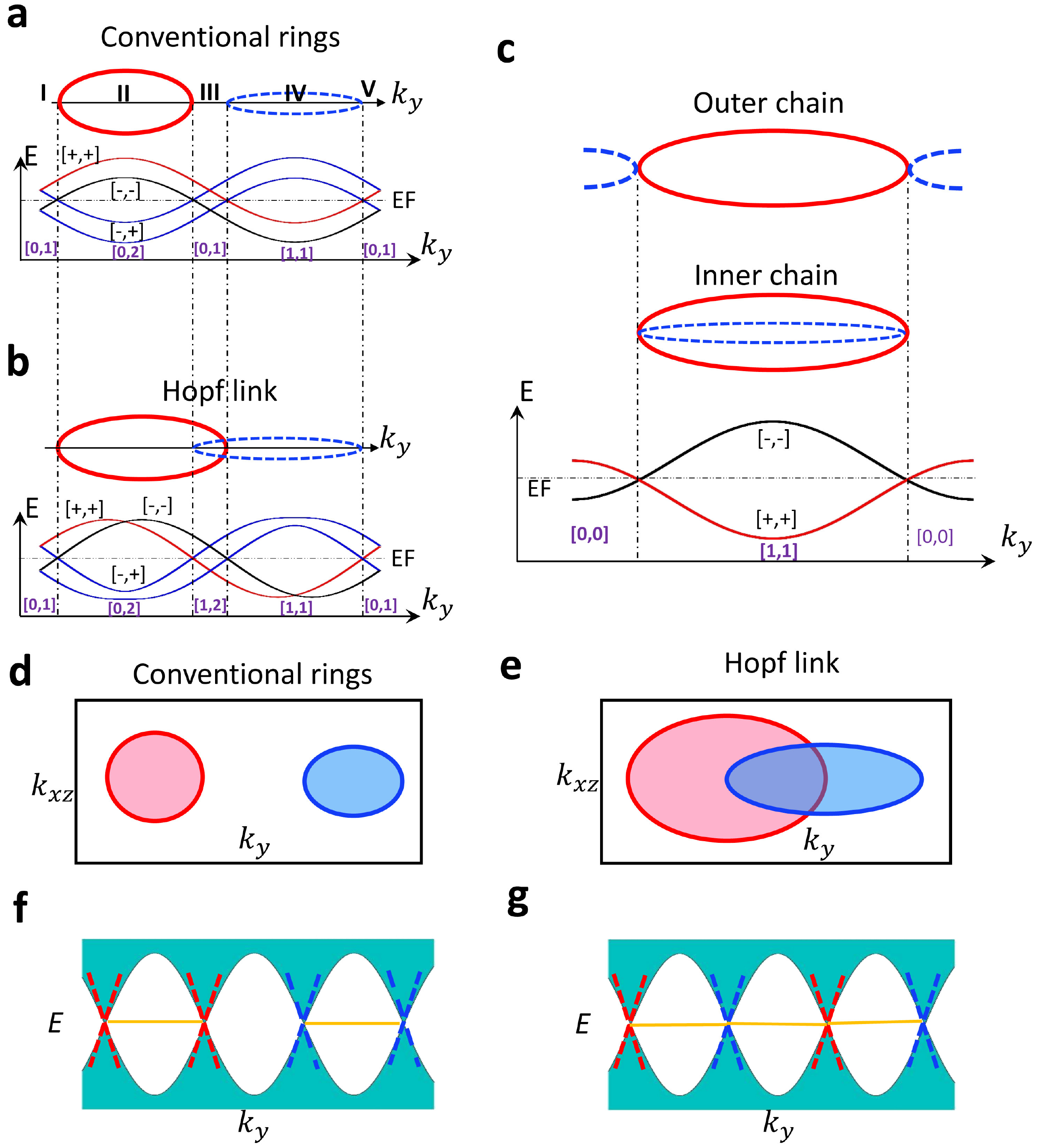}
\caption{\textbf{Mirror eigenvalues and topological surface states.} ({a}) Two independent nodal-lines (red and blue) on two perpendicular mirror planes ($M_{z}$ and $M_{x}$). The panel below shows the mirror eigenvalues ([$M_{x}$,$M_{z}$]) of the VBs and CBs along the $k_{y}$ axis. We also define a set of $\mathcal{Z}$ numbers $[\mathcal{Z}_x, \mathcal{Z}_z]$ for the two VBs. $\mathcal{Z}_x$ ($\mathcal{Z}_z$) describes the number of VBs with a positive mirror eigenvalue ($+$) on the $\mathcal{M}_x$ ($\mathcal{M}_z$) plane. ({b,c}) Same as (a) but for a Hopf link (b) and an outer/inner nodal chain (c). ({d,e}) Surface projections of two independent nodal-lines (d) and a Hopf link (e). ({f}) Independent topological drum-head topological surface states of the two independent nodal-lines. ({g}) Coupled topological drum-head surface states of a Hopf link.}
\label{Fig2}
\end{figure}

\begin{figure}
\includegraphics[width=88mm]{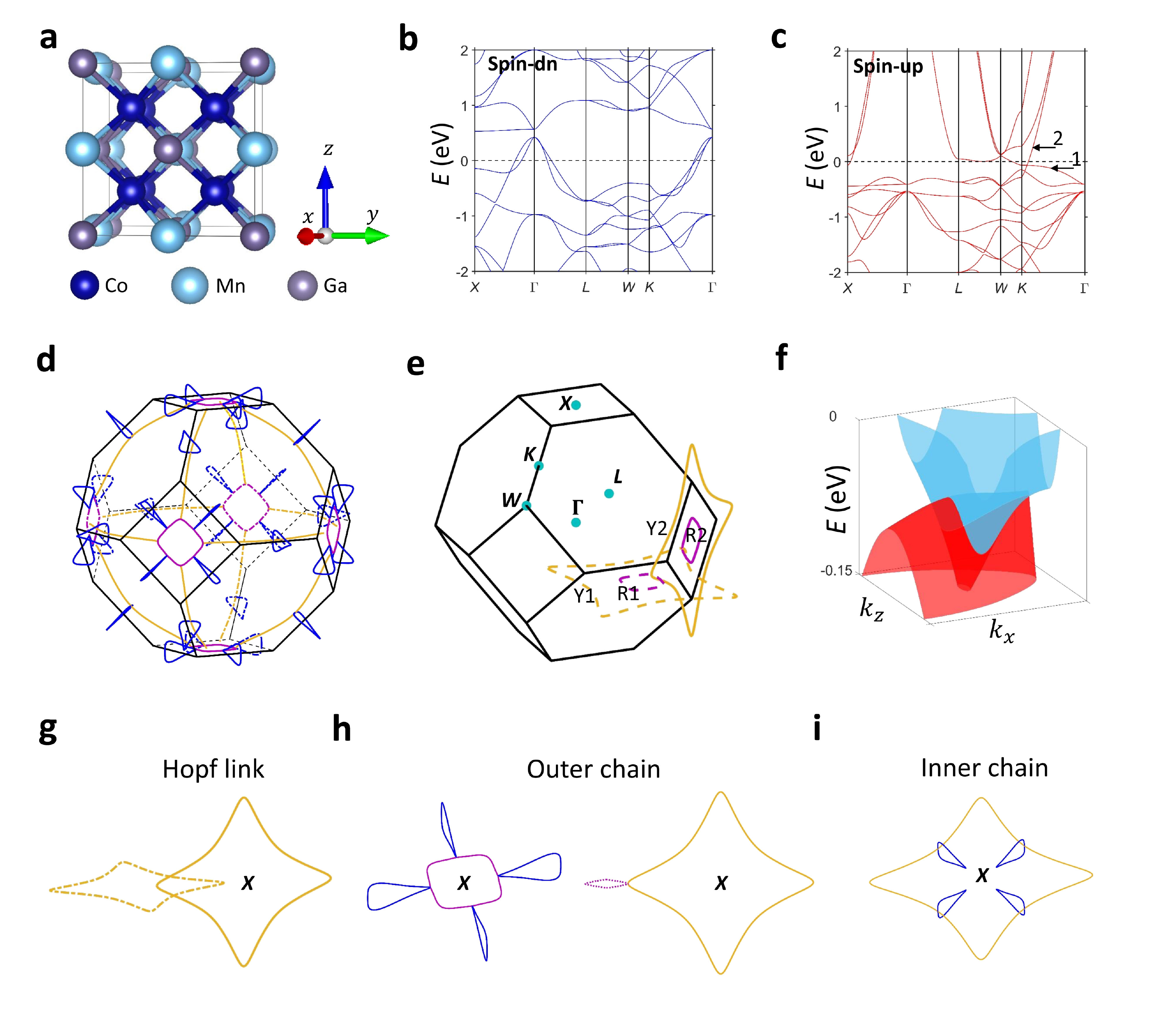}
\caption{ \textbf{Hopf links and nodal chains in Co$_{2}$MnGa.} ({a}) Crystal structure of Co$_{2}$MnGa. ({b}) Spin-down band structure. ({c}) Spin-up band structure. A band crossing is observed near the $K$ point. ({d}) Band crossings between band 1 and 2 all over the bulk BZ. Three types of nodal-lines are observed, which are colored by blue, red and yellow, respectively. The yellow and red nodal nodal-lines lie on the $M_{x}$, $M_{y}$ and $M_{z}$ mirror planes. The blue nodal nodal-lines lie on the $M_{xy}$, $M_{x\bar{y}}$, $M_{yz}$, $M_{y\bar{z}}$, $M_{xz}$ and $M_{x\bar{z}}$ mirror planes. ({e}) The red and blue nodal nodal-lines at the corner of the Brillouin zone. Y1 and R1 (Y2 and R2) are located on the $M_{z}$ ($M_{x}$) mirror plane. ({f}) 3D band structure ($E-k_x-k_z$) of a blue nodal-line. ({g-i}) Decomposing the complicated nodal features into the basis units in Fig. 1. ({g}) Two yellow nodal-lines form a Hopf link. ({h}) A blue nodal-line and a red nodal-line (or a yellow and a red nodal-line) form a outer nodal chain. ({i}) A blue nodal-line and a yellow nodal-line form an inner nodal chain.}
 \label{Fig3}
\end{figure}

\begin{figure}
\includegraphics[width=88mm]{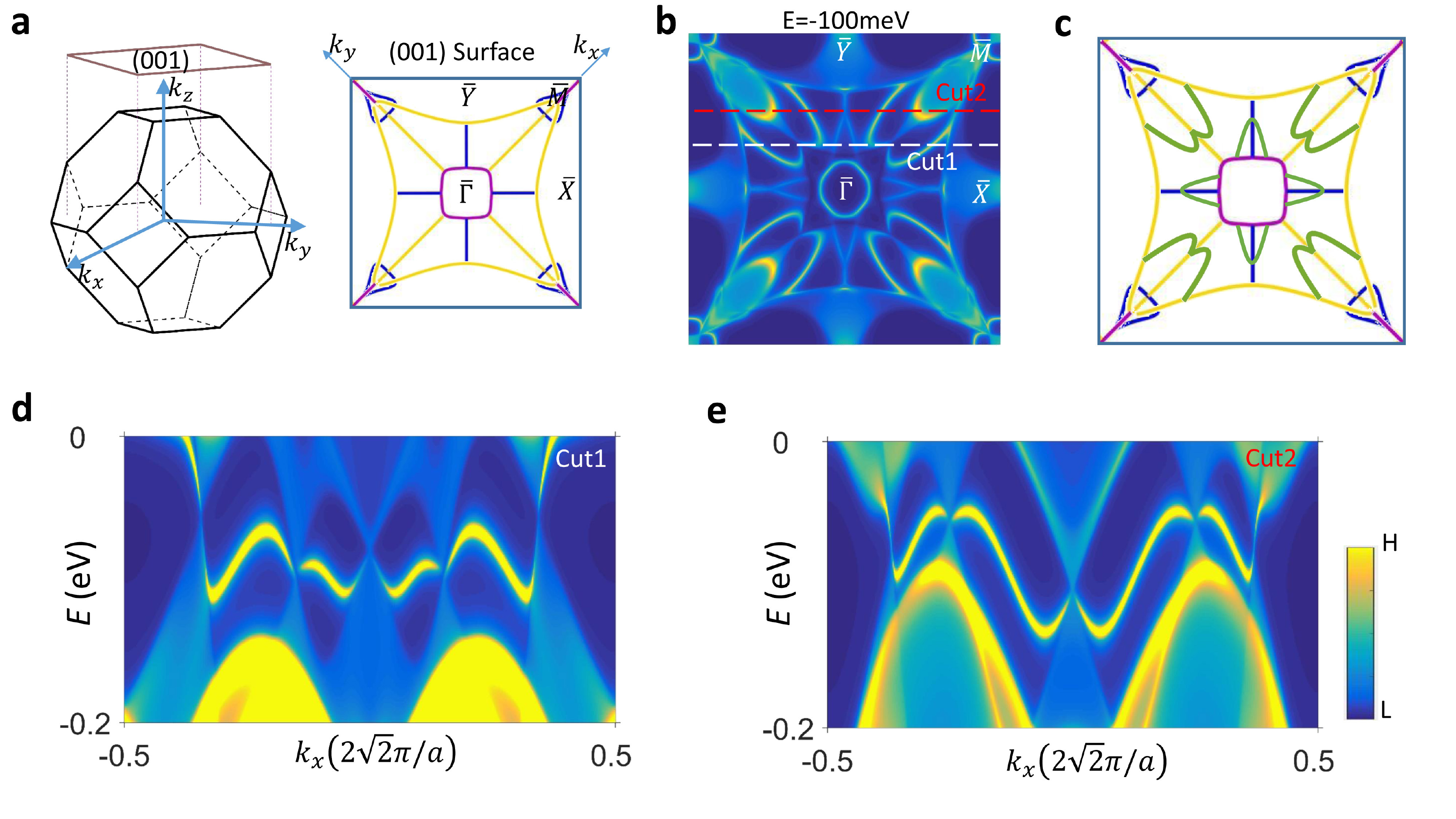}
\caption{ {Drum-head topological surface states in Co$_{2}$MnGa.} ({a}) The bulk Brillouin zone and the projection of all nodal-features onto the (001) surface.  ({b}) Surface states with Co termination at $100$ meV below the Fermi level. The sharp features are surface states whereas the shaded region are projections of bulk bands. ({c}) Schematic illustration of the surface state connectivity observed in panel (b). The green curves represent the surface states whereas other lines are the projections of the bulk nodal-lines as in panel (a). ({d, e}) Energy dispersions along the white and red cuts indicated in panel (c). Surface states connecting adjacent band crossings are observed.}
\label{Fig4}
\end{figure}

We start by considering a time-reversal broken system with one mirror plane. The breaking of time-reversal symmetry makes all bands singly degenerate. If two inverted bands have opposite mirror eigenvalues, then their crossings on the mirror plane would be protected. This leads to a mirror-protected nodal-line, as proposed in HgCr$_2$Se$_4$ \cite{Chen}, PbTaSe$_2$ \cite{PTS} and ZrPtGe \cite{ZPG}. For two perpendicular mirror planes, the normal case would be two unrelated nodal-lines on the two mirrors (Fig.~\ref{Fig1}(a)). However, there are more exotic cases (Figs.~\ref{Fig1}(b-d)). Specifically, a nodal chain consists of two perpendicular nodal-lines that touch only at a point (Figs.~\ref{Fig1}(c,d)). A Hopf link consists of two perpendicular nodal-lines that hook each other. %Below, we show how these nontrivial links arise and how they are topologically distinct from two independent nodal-lines.

In order to understand the symmetries and topological properties of these nontrivial links, we found it is helpful to study the line at which the two mirror planes coincide, i.e., the $k_y$ axis in our schematics. We consider the situation in the absence of SOC. In this case, the states are segregated into the majority spin and the minority spin that do not hybridize with each other. Only time-reversal symmetry can translate between a majority spin state and a minority spin state. By contrast, spatial symmetries (space inversion, mirror, etc) are not related to the spins without SOC. Therefore, for the states of one spin (majority or minority), all crystalline symmetries are preserved \cite{Heusler_Wang}. The effect of SOC in the real material will be discussed later. Theoretical analysis indicates that the Hopf link proteced by the mirror symmetry requires four bands. On the $k_y$ axis, these bands are described by a pair of indices, i.e., the eigenvalues of $\mathcal{M}_x$ and $\mathcal{M}_z$. As we scan along the $k_y$ axis (Figs.~\ref{Fig2}(a,b)), we expect to encounter four crossings at the Fermi level. These four crossings divide the $k_y$ axis into five regions, I-V. Regions I and V are technically the same region due periodicity. Starting from region I, we assume that the four bands have mirror eigenvalues of $[+,+]$ (red), $[-,+]$ (blue), $[-,-]$ (black), and $[-,+]$ (blue) respectively, as one decreases energy. As we go from region I to region II, we encounter the first band crossing, where the $\mathcal{M}_x$ eigenvalue remains the same whereas the $\mathcal{M}_z$ eigenvalue flips. Another way to describe the same phenomena is to define a set of $\mathcal{Z}$ numbers $[\mathcal{Z}_x, \mathcal{Z}_z]$ for the two VBs, which describes the number of VBs with a positive mirror eigenvalue ($+$) on the $\mathcal{M}_x$ ($\mathcal{M}_z$) plane. For example, in region I, the two VBs are $[-,-]$ (black) and $[-,+]$ (blue). Therefore, the $\mathcal{Z}$ numbers are $[0,1]$. In region II, the $\mathcal{Z}$ numbers become $[0,2]$. Following this construction, we obtain the $\mathcal{Z}$ numbers of the two VBs at other regions, as labeled by the purple indices in Fig.~\ref{Fig2}(a). Notably, for the nontrivial Hopf link (Fig.~\ref{Fig2}(b)), while the $\mathcal{Z}$ numbers in regions I, II, IV, and V are identical to the case of two independent nodal-lines (Fig.~\ref{Fig2}(a)), region III is opposite. The different $\mathcal{Z}$ numbers in region III demonstrate that the two independent nodal-lines (Fig.~\ref{Fig2}(a)) and the Hopf link are topologically distinct. In order to change from one to the other, one needs to invert both $\mathcal{M}_x$ and $\mathcal{M}_z$ eigenvalues between the CBs and VBs in this region.

In contrast to the Hopf link, our nodal chains on perpendicular mirror planes involve only two bands. As shown in Fig.~\ref{Fig2}(c), two band crossings divide the $k_y$ axis into two regions under a periodic boundary condition. Interestingly, for nodal chains, both mirror eigenvalues ($\mathcal{M}_x$ and $\mathcal{M}_z$) flip from one region to the other. As a result, both $\mathcal{Z}$ numbers change (e.g., from $[0,0]$ to $[1,1]$ in Fig.~\ref{Fig2}(c)). In a Hopf link, only one $\mathcal{Z}$ number changes from one region to the adjacent one. This difference highlights the topological distinction between a Hopf link and a nodal chain on perpendicular mirror planes.

We now consider the topological surface states. For a nodal-line, its topological surface states have been established as the so-called drum-head surface states that fill the entire area inside the nodal-line. This can be demonstrated by the nontrivial winding number. Essentially, the Berry phase of a $k$-space loop is $\pi$($0$) if the loop does (does not) enclose the nodal-line (Figs.~\ref{Fig2}(d,f)). By contrast, for a Hopf link, the surface projections of the two inter-linked nodal-lines overlap with each other (Fig.~\ref{Fig2}(e)). The nontrivial winding number guarantees that each band crossing must be connected to a surface state. The topological surface states fill not only the areas owned by one of the nodal-lines but also the area shared between the two nodal-lines (Fig.~\ref{Fig2}(e)). As a result, the two drumhead surface states are coupled together (Figs.~\ref{Fig2}(e,g)). We emphasize that, in contrast to two independent nodal-lines, the two nodal-lines of a Hopf link must overlap when projected on the surface. As a result, a Hopf-link's topological surface states always stitch the two interlinked nodal-lines together.

We now present Co$_2$MnGa as a material candidate of the new topological semimetal state with nontrivial links. The full Heusler Co$_2$MnGa crystal structure has a face-centered cubic Bravais lattice, with space group $Fm\bar{3}m$ (No. $225$) \cite{Heusler_crystal}, and can be viewed as consisting of exactly four interpenetrating  face-centered cubic lattices (Fig.~\ref{Fig3}(a)). This crystal has many mirror planes including $\mathcal{M}_x$, $\mathcal{M}_y$, $\mathcal{M}_z$, $\mathcal{M}_{xy}$, $\mathcal{M}_{x\bar{y}}$, $\mathcal{M}_{xz}$, $\mathcal{M}_{x\bar{z}}$, $\mathcal{M}_{yz}$, and $\mathcal{M}_{y\bar{z}}$. Here the subscript index represents the out-of-plane direction of the mirror plane. $xy$ means the direction that is $45^{\circ}$ between $\hat{x}$ and $\hat{y}$. The Co$_2$MnGa samples are known to have a high Curie temperature ($\sim700 K$) and a collinear ferromagnetic order \cite{Heusler_Curie}. Figures~\ref{Fig3}(b,c) show the bulk band structure of the spin-up and spin-down states, respectively. Near the Fermi level, the spin-dn states contribute a large hole-like pocket at the BZ center $\Gamma$ point. This pocket is relatively uninteresting from the topology point of view. We focus on the spin-ip states, which show interesting band crossings near the BZ boundaries between the lowest lying CB and VB (bands 1 and 2 in Fig.~\ref{Fig3}(c)). Figure~\ref{Fig3}(d) depicts all the band crossings between bands 1 and 2 in the bulk BZ. We found three kinds of nodal-lines, which are shown in red, yellow and blue respectively. Both the yellow and blue nodal-lines are protected by the $\mathcal{M}_x$, $\mathcal{M}_y$ and $\mathcal{M}_z$ mirror planes whereas the blue ones are protected by $\mathcal{M}_{xy}$, $\mathcal{M}_{x\bar{y}}$, $\mathcal{M}_{xz}$, $\mathcal{M}_{x\bar{z}}$, $\mathcal{M}_{yz}$, and $\mathcal{M}_{y\bar{z}}$. Remarkably, all these nodal-lines are interconnected. We decompose such a complex nodal network into the basic units described in Fig.~\ref{Fig1}. As shown in Figs.~\ref{Fig3}(g,e), two yellow nodal-lines are interlinked, forming a Hopf link. Each yellow nodal-line is also connected to four red ones to form an outer nodal chain (Figs.~\ref{Fig3}(h,e)). Moreover, each red nodal-line is further connected to four blue ones to form an outer nodal chain. Furthermore, each yellow nodal-line is connected to four blue ones to form an inner nodal chain. Therefore, all nontrivial links described in Figs. 1 and 2 are realized in Co$_2$MnGa. We have built a 6-bands tight-binding model that reproduces all the band crossings in Co$_2$MnGa. The detailed model is shown in SM \cite{SM}. We have further calculated the band structure with an on-site Coulomb repulsion $U$. We found that the band crossing is robust against on-site Coulomb repulsion $U$ while its energy may shift with respect to the Fermi level (Fig. S5). In the main text, we chose $U=0$ as the calculated magnetic moment for $U=0$ matches the best with the experimentally observed value \cite{Heusler_Curie}. Now we consider the effects of spin-orbit coupling. The inclusion of SOC couples the two spins and allows them to hybridize. Therefore, the degree of mirror symmetry breaking due to SOC can be directly measured by the SOC gap opened at the band crossings. From first principles calculations, we found that the SOC gap opened at the nodal links in Co$_2$MnGa is extremely small ($< 1$ meV, see Fig. S4). Therefore, the SOC effect is quite small. In other words, the effect of mirror symmetry breaking due to spin orientations is negligible. Similar phenomena have also been found in other full Heusler compounds \cite{Heusler_Wang, Heusler_Chang}.

Finally, we show the calculated surface band structure of Co$_2$MnGa. We observe an $\omega$-shaped surface state at the midpoint along each $\bar{\Gamma}-\bar{M}$ and a V-shaped surface state along each $\bar{\Gamma}-\bar{X}$ line. As we travel across the surface BZ along the red (white) dotted line across the surface BZ, we expect to cross the nodal-lines five times. Indeed, we found five band crossings in Figs.~\ref{Fig4}(d,e). Importantly, all surface states are found to connect the adjacent band crossings. Based on Figs.~\ref{Fig4}(b,d,e), we conclude that the $\omega$-shaped surface states connects two yellow nodal-lines (in Fig.~\ref{Fig4}b, one straight yellow line along $45$ deg and the other curved yellow line that form a large closed loop), while the  V-shaped surface state connect a blue nodal-line with the red nodal-line at $\bar{\Gamma}$. The two yellow nodal-lines form a Hopf link (Fig.~\ref{Fig3}(g)). The blue and red nodal-lines form an outer nodal chain (Fig.~\ref{Fig3}(h)). The $\omega$-shaped and the V-shaped surface states are the topological surface states. Topological surface states connecting other nontrivial links can also be identified at other binding energies.

The work at Princeton is supported by the National Science Foundation, Division of Materials Research, under Grants No. NSF-DMR-1507585 and No. NSF-DMR-1006492 and by the Gordon and Betty Moore Foundation through the EPIQS program Grant \#GBMF4547. The work at the National University of Singapore was supported by the National Research Foundation, Prime Minister's Office, Singapore under its NRF fellowship (NRF Award No. NRF-NRFF2013-03). The work at Northeastern University was supported by the US Department of Energy (DOE), Office of Science, Basic Energy Sciences grant number DE-FG02-07ER46352, and benefited from Northeastern University's Advanced Scientific Computation Center (ASCC) and the NERSC supercomputing center through DOE grant number DE-AC02- 05CH11231. The work at the National Sun Yat-sen University was supported by the Ministry of Science and Technology in Taiwan under Grant No. MOST105-2112-M110-014-MY3.

\bigskip
\textit{Note added} While finalizing our manuscript, we noticed two independent works \cite{Hopf1, Hopf2} that also proposed the Hopf-link concept. Our work is unique and distinct in two ways: (1) We consider the Hopf-link and nodal-chains enabled by perpendicular mirror plane mechanism; (2) Crucially, we also identify a real material candidate which will open the experimental research front for Hopf-link topology.

Corresponding authors:
\textbf{suyangxu@princeton.edu;
nilnish@gmail.com;
mzhasan@princeton.edu
}

\end{document}